# Compact and Energy-Efficient Memristive Spiking Neuromorphic Accelerator for Bio-inspired Interception Tasks

Qianhou Qu

Department of Electrical Engineering, The University of Texas at Arlington, qxq9423@mavs.uta.edu

Sheng Lu

Department of Electrical Engineering, The University of Texas at Arlington, sxl2408@mavs.uta.edu

Sungyong Jung

Department of Electrical Engineering and Computer Science, South Dakota State University, sungyong.jung@sdstate.edu

Qilian Liang

Department of Electrical Engineering, The University of Texas at Arlington, liang@uta.edu

Chenyun Pan

Department of Electrical Engineering, The University of Texas at Arlington, chenyun.pan@uta.edu

Spiking neural networks (SNNs) provide an efficient event-driven computing paradigm for bio-inspired interception tasks. However, most implementations rely on von Neumann digital computing platforms, where memory and computation bottlenecks limit energy efficiency. This work presents a compact and energy-efficient memristive neuromorphic accelerator for bio-inspired interception tasks. A novel one-transistor–one-resistor (1T1R) crossbar array is designed to emulate synaptic operations in the in-memory computing (IMC) domain, while circuit-level optimization mitigates membrane drift and improves integration fidelity. In addition, an integrate-and-fire (IF) neuron with separated input and membrane nodes is developed to improve inference robustness during array-interfaced operation. Implemented in the SkyWater SKY130 PDK, the proposed neuron achieves an energy consumption of 10.67 pJ/spike and an area of 906 μm². System-level results show that the memristive IMC output closely matches the software SNN baseline, with a correlation coefficient of 0.9622, while achieving a 96% interception success rate. These results demonstrate the effectiveness of the proposed design for compact and reliable memristive SNN inference in bio-inspired interception tasks.

## 1 INTRODUCTION

The brain is widely regarded as an extremely energy-efficient information processing system [1]. Information is processed through sparse and event-driven neural spikes exchanged among interconnected neurons and synapses, enabling highly parallel computation with low energy consumption. Inspired by this biological mechanism, spiking neural networks (SNNs) encode and transmit information through discrete spike events, offering significant potential for low-power and low-latency inference [2, 3]. Owing to their event-driven nature and inherent capability for temporal information processing, SNNs have emerged as a promising computing paradigm for real-time and bio-inspired applications.

Despite their biological plausibility and computational advantages, many existing SNN tasks are still executed on graphics processing unit (GPU) and central processing unit (CPU) based on the conventional von Neumann architecture [4-6]. In contrast to biological neural systems, where synaptic weights are stored locally and computation is activated only when spikes occur, these platforms separate memory and computation units, resulting in frequent data movement between them. For SNN workloads, this mismatch largely reduces energy efficiency because repeated memory accesses and dense multiply-and-accumulate operations introduce substantial data-movement overhead [7, 8], limiting their practical deployment in edge devices and real-time systems. Therefore, developing hardware architectures that process information in a brain-like, event-driven manner is critical for reducing energy consumption and improving computational efficiency.

Recently, numerous research efforts have been devoted to the development of neuromorphic computing accelerators for event-driven processing and spike-based computation. To support energy-efficient SNN inference, digital neuromorphic accelerators, such as TrueNorth [9] and Loihi [10], have demonstrated scalable and efficient spike-based computing. These platforms provide flexible programmability and large-scale neural integration, making them important milestones in neuromorphic hardware development. However, digital neuromorphic accelerators still fundamentally rely on separated memory and computation resources for synaptic processing, while clocked digital control limits efficiency and natural support for continuous-time neural dynamics.

In addition to digital neuromorphic platforms, analog neuron circuits and analog in-memory computing (IMC) structures have been widely explored for implementing brain-inspired spike-based processing [11-17]. By exploiting continuous-time circuit dynamics, these neuron designs can naturally realize membrane integration, threshold comparison, and spike generation in an asynchronous and energy-efficient manner [12, 13, 15, 17]. In these prior works, the membrane potential is accumulated directly on a capacitor connected to the input node, such that incoming synaptic currents charge the membrane capacitor to perform temporal integration. Although this approach is effective for standalone neuron operation, its accuracy can be affected when the neuron is directly interfaced with array-based IMC systems. One way to alleviate this issue is to introduce peripheral analog-to-digital converters (ADCs) and digital-to-analog converters (DACs) to digitize the array output before subsequent processing. However, such a solution introduces considerable area, energy, and latency overhead, reducing the compactness and efficiency advantages of analog IMC hardware [18, 19]. Thus, it is important to design a neuron circuit that enables accurate temporal integration with minimum hardware cost.

On the other hand, efficient hardware realization of SNN inference also critically depends on the design of synaptic circuits and arrays. In SNN accelerators, synapses are responsible for storing connection weights and performing weighted accumulation of input spikes. Conventional circuit-based synapse implementations often require multiple transistors, which introduce area overhead when scaled to large arrays. To address this challenge, memristive devices have attracted considerable attention for synaptic implementation due to their nonvolatile storage capability, analog programmability, and high integration density. In several prior works [8, 20-22], memristive crossbar arrays have been employed to perform analog weighted accumulation with high parallelism and reduced data movement, thereby improving the area and energy efficiency of synaptic computation. However, most of these studies focus primarily on the mapping of synaptic weights and the implementation of crossbar-based accumulation, while the impact of read noise during inference is often left underexplored. In practical memristive arrays, read noise can perturb the effective conductance values and accumulated current, which may further affect spike generation and degrade the inference accuracy of SNN accelerators.

In this paper, we introduce a novel neuron circuit that decouples the membrane-potential storage node from the input port, substantially improving temporal-integration fidelity, stabilizing spike generation, and avoiding costly peripheral data-conversion overhead. We employ a one-transistor-one-resistor (1T1R) crossbar structure to enable compact synaptic weighted accumulation directly in the IMC domain, while explicitly incorporating memristive read nonidealities to preserve reliable hardware inference under realistic operating conditions. Together, these design choices enable a compact, robust, and energy-efficient SNN accelerator for bio-inspired interception tasks. For the rest of the paper, Section 2 first establishes the bio-inspired interception task formulation, the corresponding SNN framework, and the training strategy that serves as the algorithmic basis of this work. Section 3 then presents the proposed memristive neuromorphic accelerator and explains the key circuit- and device-level design considerations for neuron operation, synaptic interfacing, and weight realization in the presence of memristor nonidealities. Based on this design framework, Section 4 validates the proposed approach through circuit- and system-level results, demonstrating the inference accuracy, closed-loop task performance, robustness under variations, and implementation efficiency of the hardware. Finally, Section 5 concludes the paper.

The main contributions of this paper are summarized as follows:

- A compact and energy-efficient memristive neuromorphic accelerator is proposed for bio-inspired interception tasks, enabling spike-based inference through analog in-memory computing.
- An integrate-and-fire (IF) neuron circuit with separated input and membrane nodes is developed to mitigate the nonlinear integration and saturation behavior of conventional coupled-node neurons, thereby improving temporal-integration fidelity and spike-generation reliability in array-interfaced IMC operation.
- A read-noise-aware memristive synapse realization and hardware-mapping flow is introduced by combining closed-loop write–verify programming with two-device parallel mapping.
- A comprehensive software-hardware co-design and evaluation framework is developed to validate hardware accuracy and task-level effectiveness.

## 2 NETWORK DESIGN AND TRAINING METHODS

In this section, we first introduce the predator-prey interception task and its sensory variables, then present the SNN formulation and spike-based encoding scheme, and finally describe the supervised training methodology.

### 2.1 Predator-Prey Tracking Task

Inspired by the interception model in [23], we formulate a two-dimensional predator-prey tracking task to evaluate the proposed memristive neuromorphic hardware. In this task, the predator and the prey move in a 2-D continuous space with step-based motion. In each episode, the predator is initialized at the origin (0, 0), while the prey is placed at (8, 0). The prey moves in a 2-D continuous space with a constant step size of 1 along a stochastic trajectory, while the predator moves with a slightly higher step size of 1.3. An interception is considered successful when the distance between the predator and the prey is reduced to 1 or below.

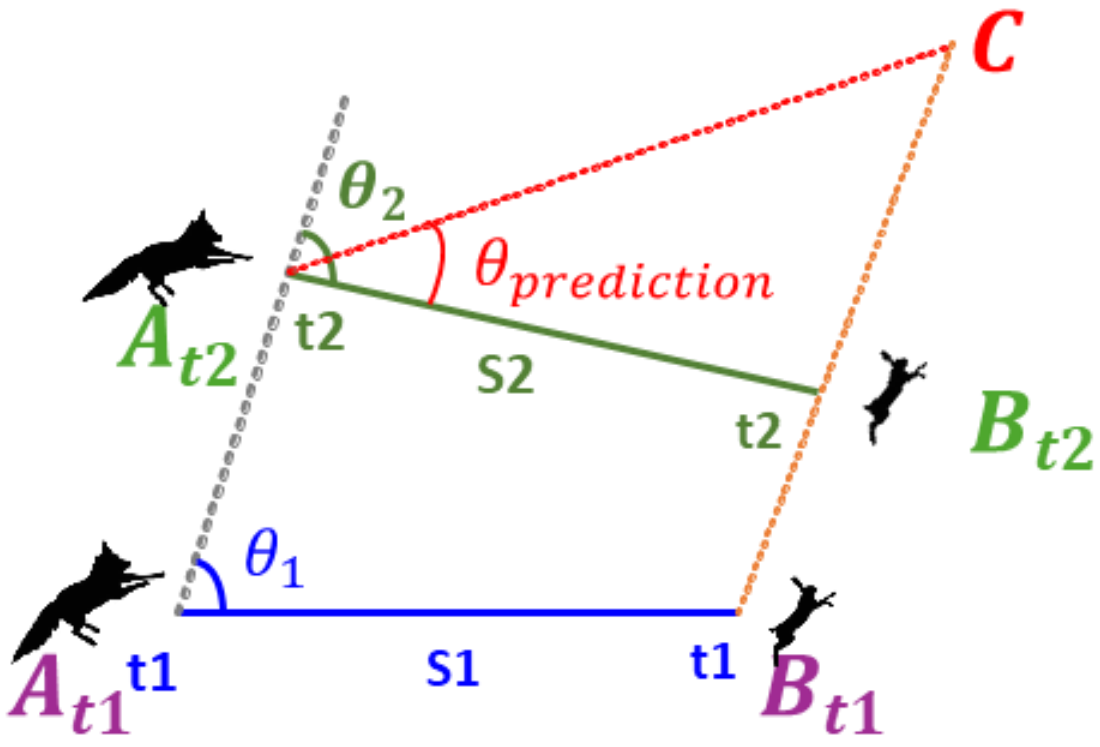

Fig. 1. SNN task representation showing relative distance and bearing angles used to predict the next turning angle.

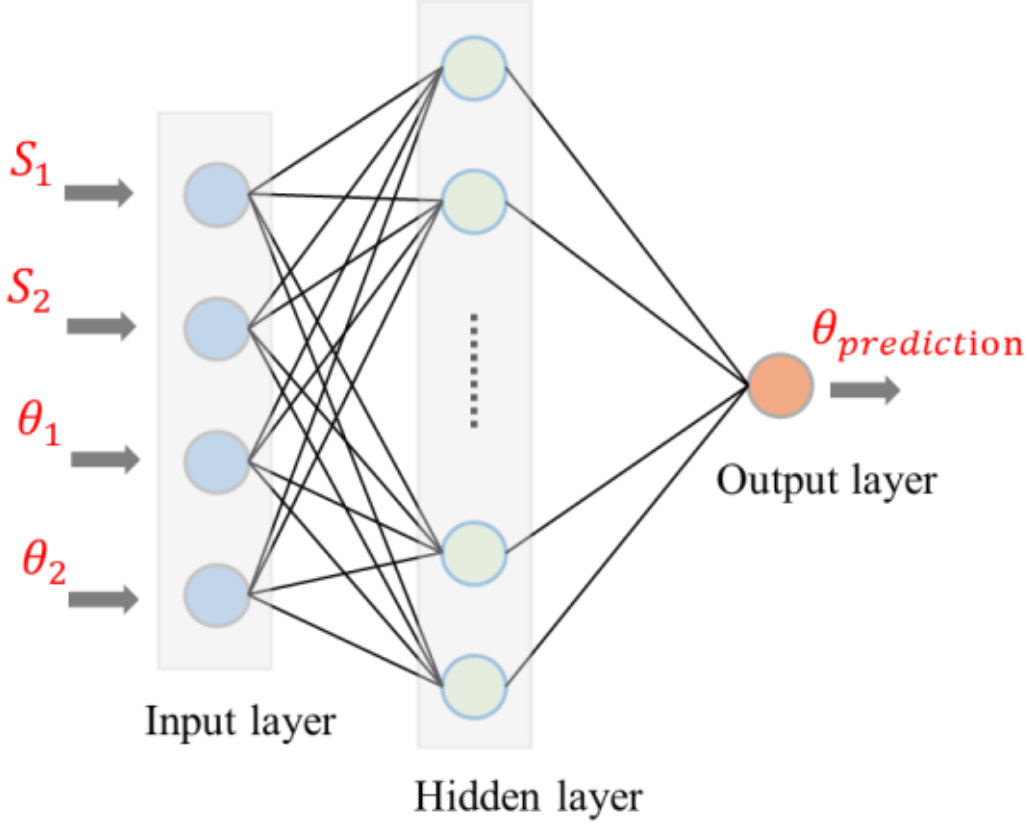

Fig. 2. SNN architecture for predator pursuit task.

To determine the predator steering direction at each step, four relative kinematic variables are extracted from the predator-prey geometry shown in Fig. 1. Specifically, S1 denotes the predator-prey distance at the previous time step, while S2 denotes the predator-prey distance at the current time step. The angle θ2 is defined as the relative angle between the predator moving direction before updating and the current line-of-sight direction from the predator to the prey. The angle θ1 is defined as the angle between the predator heading after the steering update at the previous step and the predator-to-prey line-of-sight direction at that same step. These four variables jointly represent the instantaneous interception state and are used as the inputs to the 4-30-1 SNN, as shown in Fig. 2**.** After deterministic spike encoding, the input spike trains are first projected to the hidden layer through weighted synaptic connections. The hidden IF neurons accumulate the incoming weighted spikes in their membrane states and emit output spikes when the firing threshold is reached. The hidden-layer spike outputs are then propagated to the output neuron through the second synaptic layer, where temporal integration is performed again to generate the final output spike train. By counting and normalizing the output spikes over the full encoding window, the network predicts the steering angle for the next motion step.

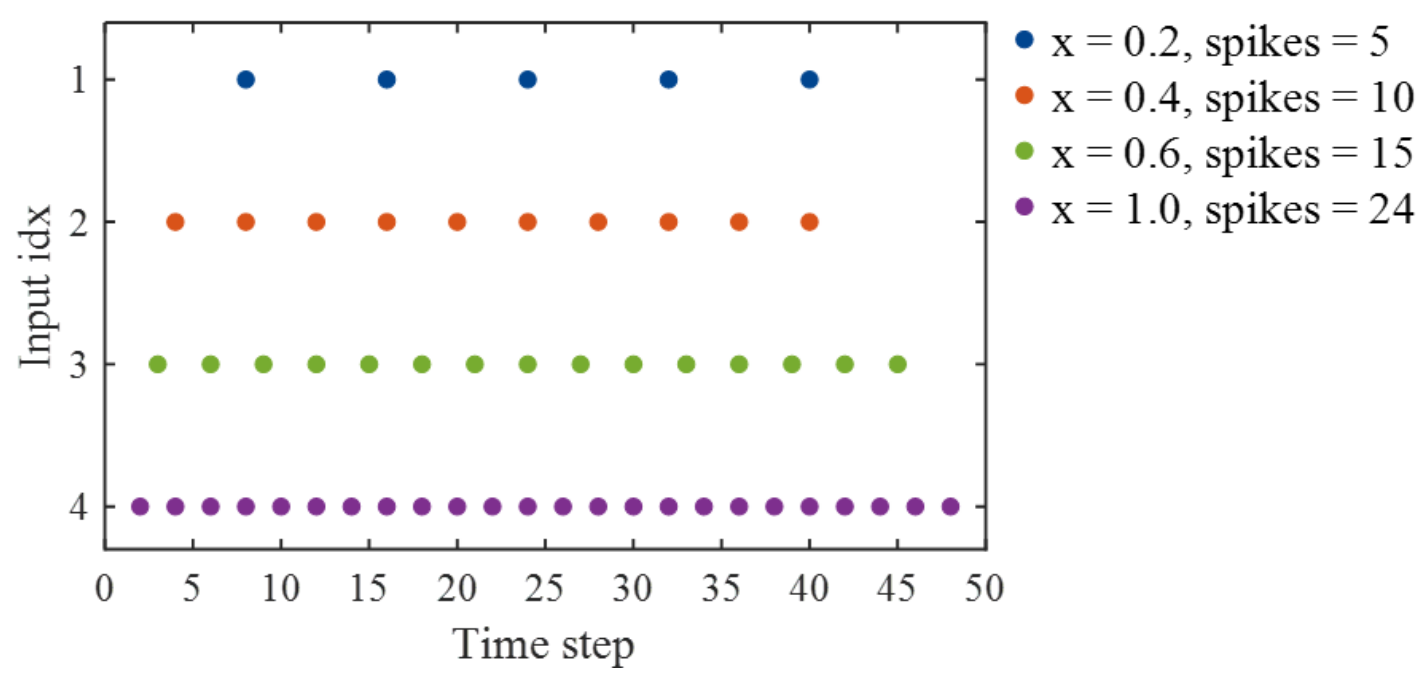

Fig. 3. Illustration of deterministic uniform rate encoding for different normalized inputs.

### 2.2 Input and Output Encoding Scheme for SNN

To convert the continuous kinematic variables into spike-based inputs suitable for SNN processing, a deterministic uniform rate-coding scheme is adopted. The four input variables, ordered as [$S_1$, $S_2$, $\theta_1$, $\theta_2$], are first normalized to the range [0,1] and then mapped into discrete spike trains over a fixed simulation window. Specifically, the two distance variables are normalized using a distance scale of 10, while the two angular variables are linearly shifted and scaled according to their predefined angle ranges before being clipped to [0, 1]. In the implementation, $\theta_1$ is mapped based on the range from -60° to 60°, and $\theta_2$ is mapped based on the range from -100° to 100°. This encoding method is used to establish a direct correspondence between the algorithm-level SNN and the pulse-based hardware implementation.

Specifically, the encoding window is set to 50 time steps, and the maximum number of spikes for each input neuron is limited to 25. For a normalized scalar input $x \in [0, 1]$, the spike count is first determined by

$$n = round(x \cdot N_{max}) \tag{1}$$

where $N_{max} = 25$ and n ranges from 0 to 25. If $n = 0$, no spike is generated. Otherwise, the n spikes are placed as uniformly as possible over the available time steps from 1 to $T - 1$, where $T = 50$. In the implementation, the interval is computed from the available time steps and rounded to an integer, and spikes are then generated periodically according to this interval. Therefore, a larger input value produces a denser spike train, while a smaller value leads to fewer spikes with larger spacing. In this work, the spike budget is limited to 25 as a tradeoff between encoding resolution and hardware efficiency. Such a moderate spike count helps reduce switching activity and inference energy in the proposed memristive IMC hardware while still providing sufficient representation capability for the target interception task. An example of the deterministic uniform rate-coding scheme for different normalized inputs is illustrated in Fig. 3.

The network output follows the same spike-based interpretation. After the output neuron generates spikes over the same simulation window, the total spike count is divided by the maximum spike count to obtain the normalized prediction. This normalized output is then used to represent the steering command of the predator for the next motion step. In this way, both the input encoding and output decoding are defined in a consistent spike-count domain, which simplifies the mapping from software SNN inference to the proposed memristive IMC hardware.

### 2.3 Network Training Samples and Methodology

Ideal training labels are generated using a geometrical interception model. Let $A_t$ and $B_t$ denote the positions of the predator and prey at time step t, respectively. For each time step, an interception point C is defined such that the arrival times of the predator and prey at C are identical. This interception condition ensures that the predator does not perform pure pursuit of the prey's instantaneous position but instead leads the prey toward a future collision point. Under this condition, the angle $\angle CAB$, defined by the predator position $A_t$, prey position $B_t$, and interception point $C$, represents the ideal steering angle that the predator should take at time step t.

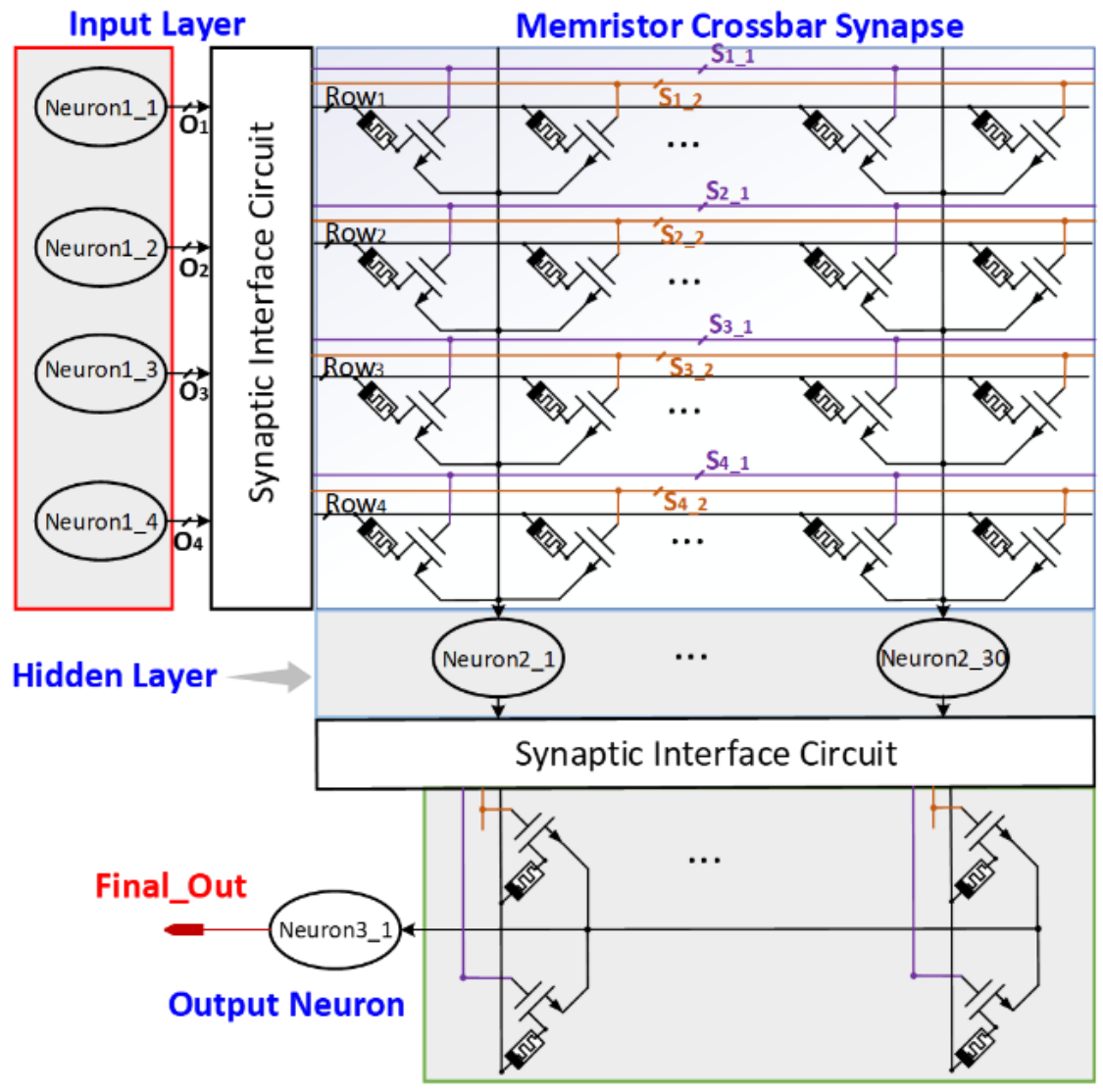

Fig. 4. Overall architecture of the proposed memristive neuromorphic accelerator.

The SNN is trained in a supervised manner using SNNTorch [24] with backpropagation through time (BPTT). During training, the encoded spike sequences are fed into the network, and the normalized output spike count is used as the predicted steering command. The mean squared error (MSE) between the predicted and ideal turning angles is adopted as the training loss, and the network parameters are optimized using the Adam optimizer. In addition, the synaptic weights are clipped after each update to improve training stability and facilitate subsequent hardware mapping. In this work, 200 predator-prey episodes are generated for training, with 200 training epochs, a batch size of 256, and a learning rate of 0.001.

## 3 PROPOSED ACCELERATOR ARCHITECTURE

In this section, we present the proposed memristive neuromorphic accelerator architecture for bio-inspired interception and analyze its key circuit- and device-level design features. We first introduce the overall hardware organization and its mapping from the trained 4-30-1 SNN. Next, the proposed IF neuron circuit is described and its integration behavior is evaluated. We then introduce the synaptic interface circuit, which connects the neuron output to the memristive crossbar array and provides the required signal conditioning and driving capability. Finally, the memristive cell implementation, multi-level weight mapping, and read-noise modeling are presented.

### 3.1 Overall System Architecture and Network Mapping

Fig. 4 shows the overall architecture of the proposed memristor-based SNN accelerator with a 4-30-1 network topology. The system consists of an input neuron layer, a hidden neuron layer, an output neuron, and the memristor crossbar synapses between adjacent layers.

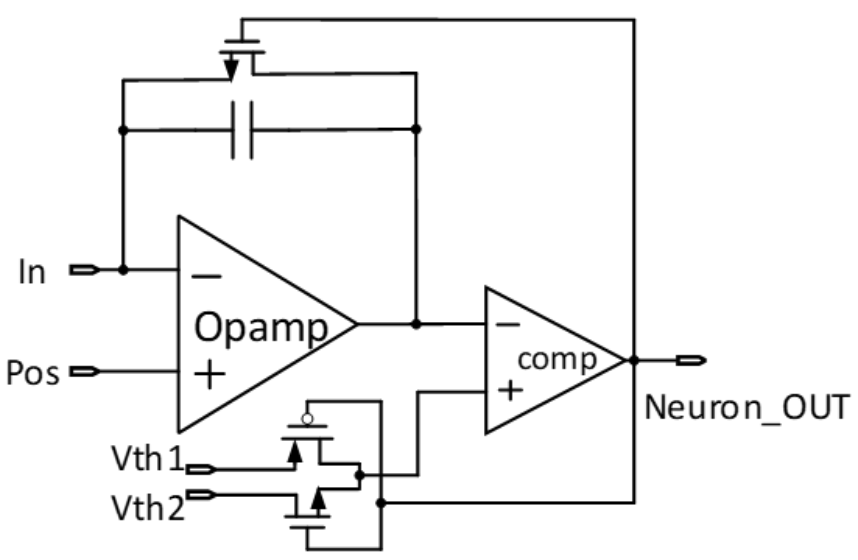

Fig. 5. Schematic of the proposed IF neuron circuit.

For each layer-to-layer connection, the spike outputs of the presynaptic neurons are first delivered to the synaptic interface circuit, which drives the row lines of the corresponding crossbar array. In the crossbar, each synaptic weight is mapped by two memristor cells, denoted as branches A and B. To enable independent control of these two branches, each row is associated with two control lines, namely the purple signal line $S_{i_1}$ and the orange signal line $S_{i_2}$. Under this organization, the weighted signals generated by the selected memristor branches are collected along the column lines and delivered to the postsynaptic neurons. Accordingly, the first crossbar stage maps the four input neurons to the 30 hidden neurons, while the second synaptic stage connects the hidden-layer outputs to the final output neuron.

### 3.2 Proposed IF Neuron Circuit and Integration Behavior

Fig. 5 shows the proposed IF neuron circuit. The neuron is built around an integrator for membrane-voltage accumulation and a comparator for threshold detection, which determines the spike generation and reset process by comparing the membrane voltage with the reference voltages.

In the proposed design, the resting potential is set to $V_{pos} = 0.9\ V$, and the membrane capacitor is $73.5\ fF$. The neuron integrates from 0.9 V downward, rather than upward. Two reference voltages, $V_{th1} = 0.71\ V$ and $V_{th2} = 0.88\ V$, are introduced to control the firing and reset operations. Specifically, $V_{th1}$ defines the firing threshold of the neuron. Once the membrane voltage drops from the resting level to 0.71 V, the comparator is triggered and the neuron generates an output spike. The second threshold, $V_{th2}$, is introduced to ensure a complete reset of the membrane voltage after firing. Without $V_{th2}$, the comparator output would return to zero as soon as the membrane voltage rises back above $V_{th1}$, which would prematurely disable the reset path and leave the membrane voltage only partially restored. By using $V_{th2}$, the reset operation is maintained until the membrane voltage is sufficiently recovered to the resting state, thereby guaranteeing stable neuron operation across consecutive spikes.

Fig. 6 shows the transient response of the proposed IF neuron under repetitive input pulses. The upper panel presents the input pulse train applied to the neuron, while the lower panel shows the corresponding membrane voltage and output spike waveform. As the input pulses arrive, the membrane voltage decreases step by step due to temporal integration. Once the membrane voltage reaches the firing threshold, the neuron generates an output spike. The neuron output is subsequently fed to the synaptic interface circuit, which performs signal conditioning for the subsequent memristive array operation. The resulting waveform after synaptic interface circuit processing is shown by the red line in Fig. 6, where a 60 ns output pulse is obtained. The detailed circuit implementation and functionality of the synaptic interface circuit will be described in Section 3.3.

To further demonstrate the advantage of the proposed design, we note that several prior works, [12, 13, 17, 25, 26] adopt neuron structures in which the membrane capacitor node is directly connected to the neuron input path. As shown in spike timing and spike order are critical to the computation flow in SNN-based IMC systems, such timing deviation can propagate across layers and ultimately degrade the overall inference accuracy.

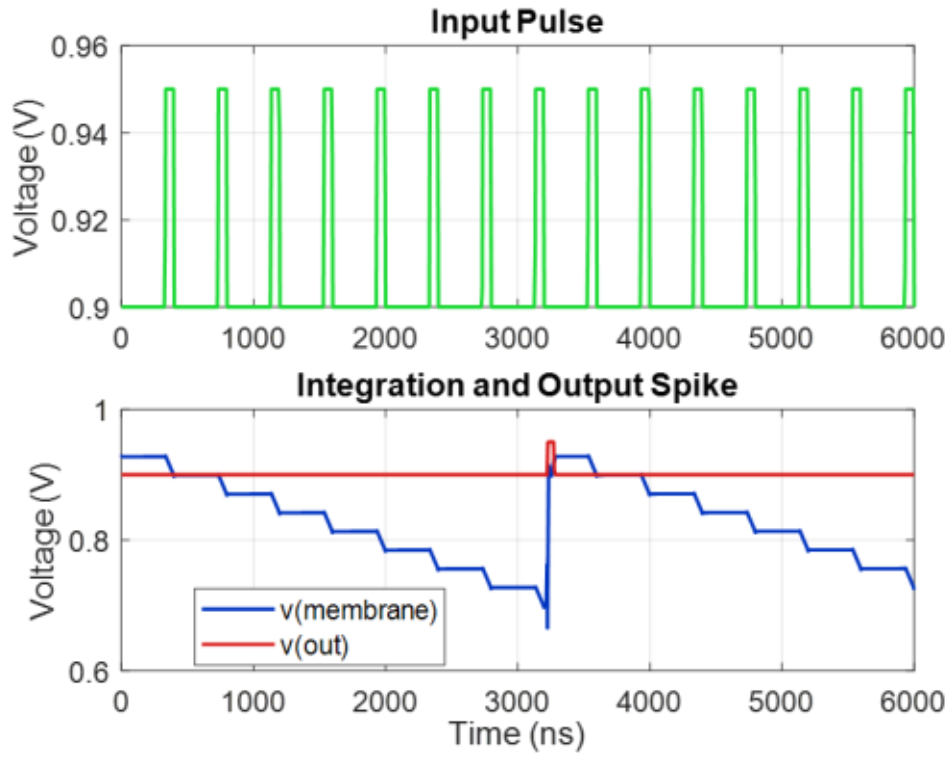

Fig. 6. Membrane voltage integration and output spike generation of the proposed IF neuron under repetitive input pulses.

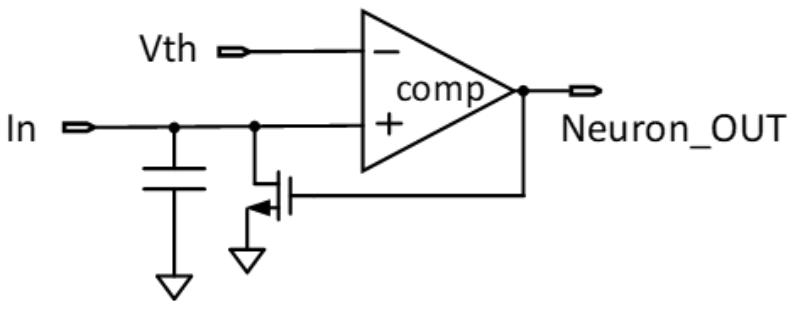

Fig. 7. Schematic of the conventional neuron with directly coupled input and membrane nodes.

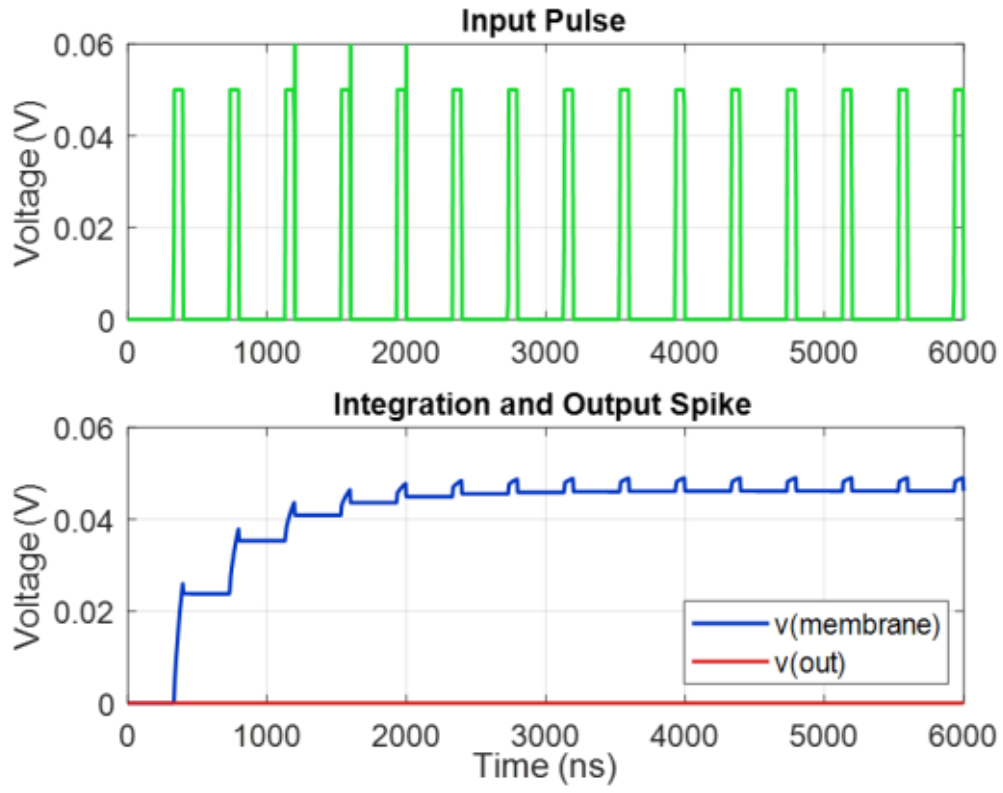

Fig. 8. Nonlinear membrane-voltage accumulation of the conventional coupled-node neuron under repetitive input pulses.

Fig. 7, a common characteristic of these structures is that the input node is directly connected to a membrane capacitor for temporal integration, while a comparator-like circuit determines whether the neuron should fire, and a transistor-based reset path is used to restore the membrane voltage after spike generation. To provide a qualitative comparison, we constructed a comparison case and applied the same integration operation. As shown in Fig. 8, the conventional neuron exhibits evident non-ideal behavior in the later stage of integration. As more input spikes arrive, the membrane-voltage increment gradually becomes saturated and even unstable, rather than maintaining the desired linear accumulation. This is mainly because the input node and the membrane node are tied together, causing the effective charging condition to change continuously as the membrane voltage rises. As a result, the timing of spike generation can be significantly distorted. Since

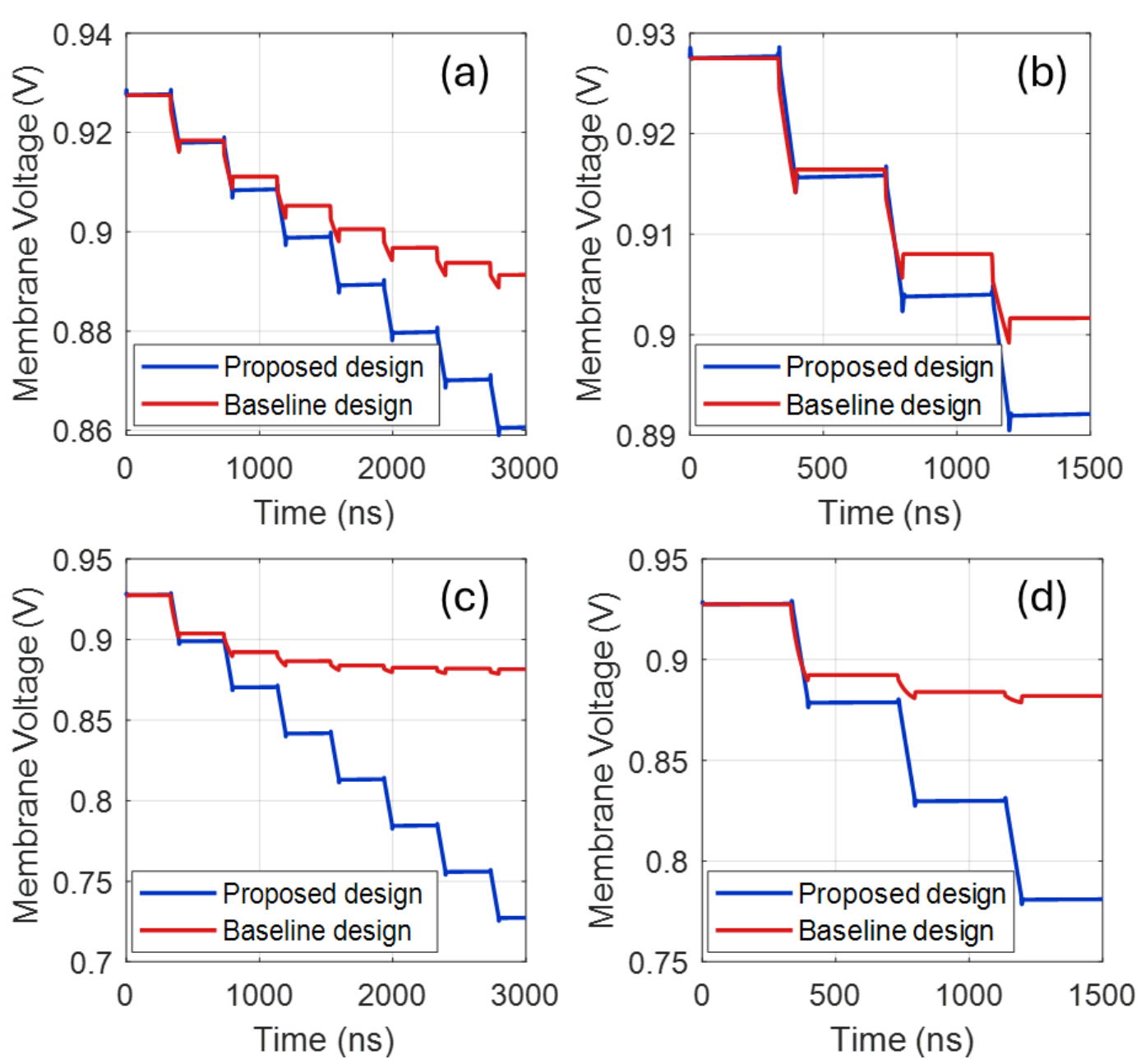

Fig. 9. Comparison of membrane voltage integration between the proposed neuron and a conventional coupled-node neuron.

Another important limitation is that the final saturation level of the membrane voltage is directly determined by the amplitude of the presynaptic output spike. In the comparison case, the presynaptic spike amplitude is 50 mV. Consequently, as observed in Fig. 9, the membrane-voltage integration saturates around 50 mV and cannot continue to increase toward the designed firing threshold, which is 0.2 V in this work. In principle, this saturation ceiling could be raised by increasing the spike amplitude. However, doing so would apply larger voltage pulses to the memristor crossbar, which may introduce undesirable write disturbance and compromise the reliability of read-domain computation. On the other hand, the neuron firing threshold may be reduced to below 50 mV so that spike generation can still be triggered under this limited integration range. However, such a low-threshold design does not eliminate the nonlinear integration behavior caused by the coupled input and membrane nodes. Moreover, to obtain a more stable integration response under this constrained operating range, a relatively larger membrane capacitor may still be required, leading to increased area overhead. In comparison, the proposed design isolates the membrane node from the input path and therefore enables more stable membrane integration and more reliable spike generation.

### 3.3 Synaptic Interface Circuit

The synaptic interface circuit is designed to bridge the neuron core and the subsequent crossbar array, while simultaneously shaping the neuron output waveform into a form suitable for the next-stage integration. As shown in Fig. 10, this interface serves two main purposes. First, it provides the driving capability required for the row-level signal propagation in the array. Second, it reshapes the neuron output pulse into a controlled voltage swing from 0.9 to 0.95 V, so that the generated signal can be directly used as the input for the postsynaptic neuron in the following stage.

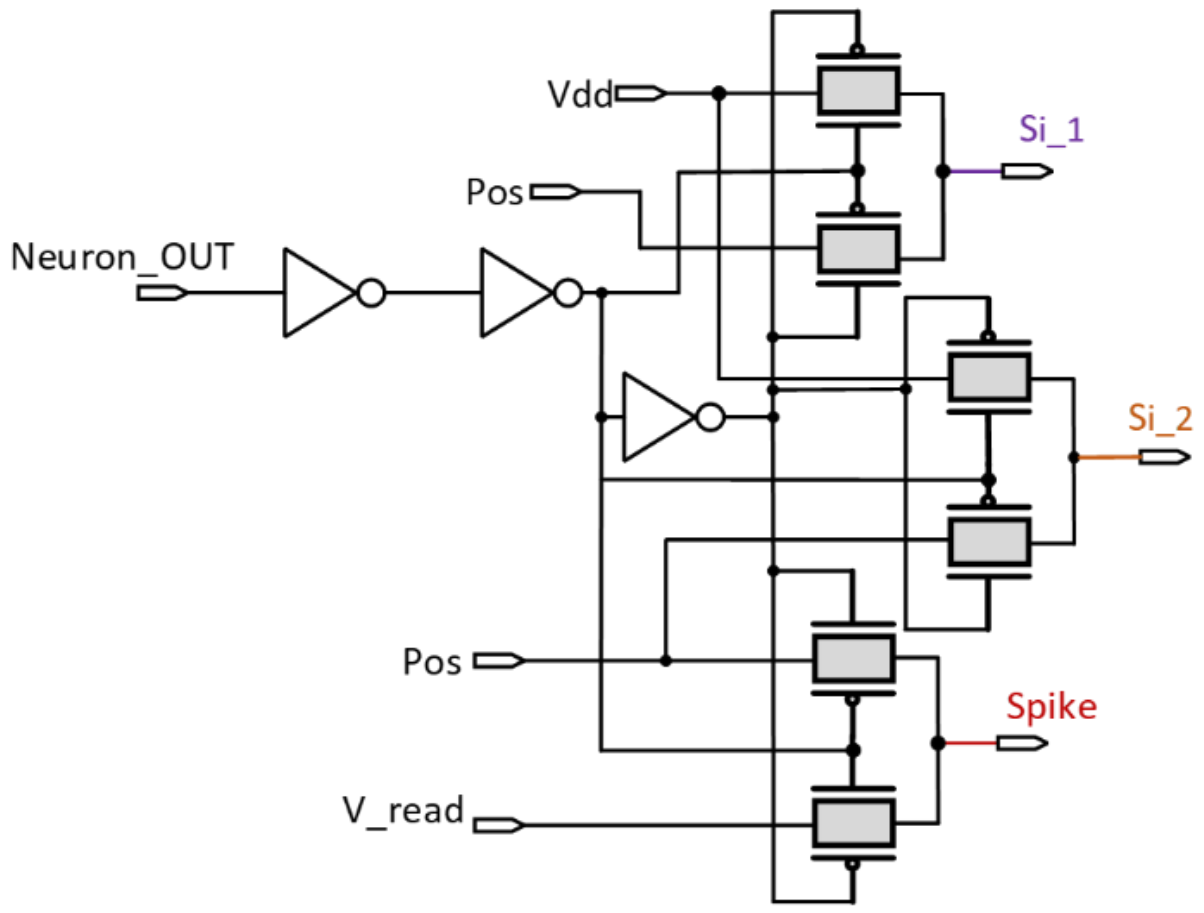

Fig. 10. Schematic of the synaptic interface circuit for row-line driving and output spike conditioning.

To achieve this goal, a buffer stage is inserted between the neuron circuit and the array-driving/output-switching circuitry. The introduction of this buffer is essential because directly connecting the neuron output to the subsequent transmission-gate network results in a noticeable pulse-width broadening. Since the transmission gates serving as the array-driving interface are designed with relatively large device sizes, they introduce considerable gate and parasitic capacitances. When such a capacitive load is directly driven by the neuron output stage, the output transition is slowed down, leading to widened pulses and degraded edge sharpness. Such pulse broadening is undesirable in the proposed system, since it may introduce unintended disturbance during the memristor read operation and increase the risk of read-induced write noise. By isolating the neuron core from the heavily loaded output network, the buffer effectively sharpens the pulse edge and reduces the output pulse width, thereby improving the reliability of signal transmission for array-level operation.

The output interface is composed of three pairs of transmission gates, which generate three coordinated control signals for the synaptic array. Among them, the signal labeled Spike is directly connected to the corresponding row line. This signal carries the presynaptic neuron activity and is used as the driving waveform applied to the selected row during the read/integration process. In addition, the signals labeled $S_{i_1}$ and $S_{i_2}$ are connected to the gate terminals of the two access nFETs associated with the two memristor branches in each cell on the same row. These two control signals are therefore used to enable or disable the corresponding current paths through the synaptic cell. When the presynaptic neuron remains in the resting state, these paths are turned off. When a firing event occurs, the interface activates the row-driving pulse and the associated switch-control signals, enabling the desired synaptic read path for postsynaptic accumulation.

### 3.4 Memristive Synapse Implementation

#### *3.4.1Memristor Model*

Memristor is a two-terminal resistive switching device whose resistance state can be modulated and retained, making it attractive for non-volatile memory and in-memory computing applications. The memristor concept was first theoretically introduced by Chua in 1971 [27] and later gained broad attention after the experimental demonstration of metal-oxide resistive switching devices in 2008 [28]. Owing to their non-volatility, compact structure, and analog programmability, memristive devices have become promising candidates for synaptic-weight implementation in neuromorphic and IMC systems.

In this work, the memristive synapse is modeled using the Peking University-Stanford RRAM compact model (v2.0Beta), which is a physics-based Verilog-A/SPICE model for metal-oxide bipolar RRAM [29, 30]. The adopted model is built upon the conductive filament (CF) evolution mechanism during SET and RESET operations. In particular, the device state is described by the evolution of the filament gap and filament width, while the electrical transport is modeled by the combination of ohmic conduction and generalized tunneling conduction. The model framework also includes parasitic components of the metal-insulator-metal (MIM) structure and self-heating effects, enabling compact yet physics-grounded representation of RRAM switching behavior.

#### *3.4.2Read Noise Modeling*

Prior experimental study [31] has shown that HfOx-based memristor exhibits noticeable current fluctuation during read operation, which is commonly characterized as random telegraph noise (RTN). In particular, the relative resistance fluctuation, $\Delta R/R$, is strongly dependent on the device resistance: it increases as the resistance increases and gradually saturates in the high-resistance region. This observation suggests that the read noise in RRAM is highly state dependent and should be incorporated into device/circuit modeling accordingly.

To capture this effect in large-scale circuit and system simulations, we incorporate an RTN-inspired effective read-noise model into the Verilog-A RRAM model. Instead of explicitly modeling trap-level capture/emission events and discrete two-level temporal switching, we focus on reproducing the experimentally observed dependence of relative read fluctuation on device conductance. In the proposed implementation, the ideal read current $I_{ideal}$ is first computed from the original device equations, and the effective read conductance is defined as

$$G_{read} = \frac{|I_{ideal}|}{|V_{ref}|} \tag{2}$$

The relative noise strength is then modeled as a conductance-dependent power law

$$\sigma_{raw} = \sigma_0 (\frac{G_{read}}{G_0})^{-p_\sigma} \tag{3}$$

where $\sigma_0$ denotes the reference relative fluctuation at conductance $G_0$, and $p_\sigma$ controls how the fluctuation scales with device conductance. In this work, the parameter values are chosen as $\sigma_0 = 0.035$, $G_0 = 7 \times 10^{-6}\ S$ and $p_\sigma = 0.95$.

To ensure numerical stability and avoid unrealistic extremes, the relative noise magnitude is further processed by a bounded smooth saturation function. Specifically, for $\sigma_{raw} \le \sigma_{min}$, the effective value is set to $\sigma_{final} = \sigma_{min}$. While for $\sigma_{raw} > \sigma_{min}$,

$$y_{sat} = \frac{\sigma_{raw} - \sigma_{min}}{\sigma_{max} - \sigma_{min}} \tag{4}$$

$$\sigma_{final} = \sigma_{min} + (\sigma_{max} - \sigma_{min}) \frac{y_{sat}}{1 + y_{sat}} \tag{5}$$

In the present model, the lower and upper bounds are set to $\sigma_{min} = 0$ and $\sigma_{max} = 0.35$.
The final noisy read current is expressed as

$$I(t,b) = I_{ideal} + I_{noise} \tag{6}$$

with

$$I_{noise} = \sigma_{final} I_{ideal} N(t) \tag{7}$$

where $N(t)$ is a Gaussian random variable updated with a sample-and-hold interval $\tau_{noise} = 10ns$.

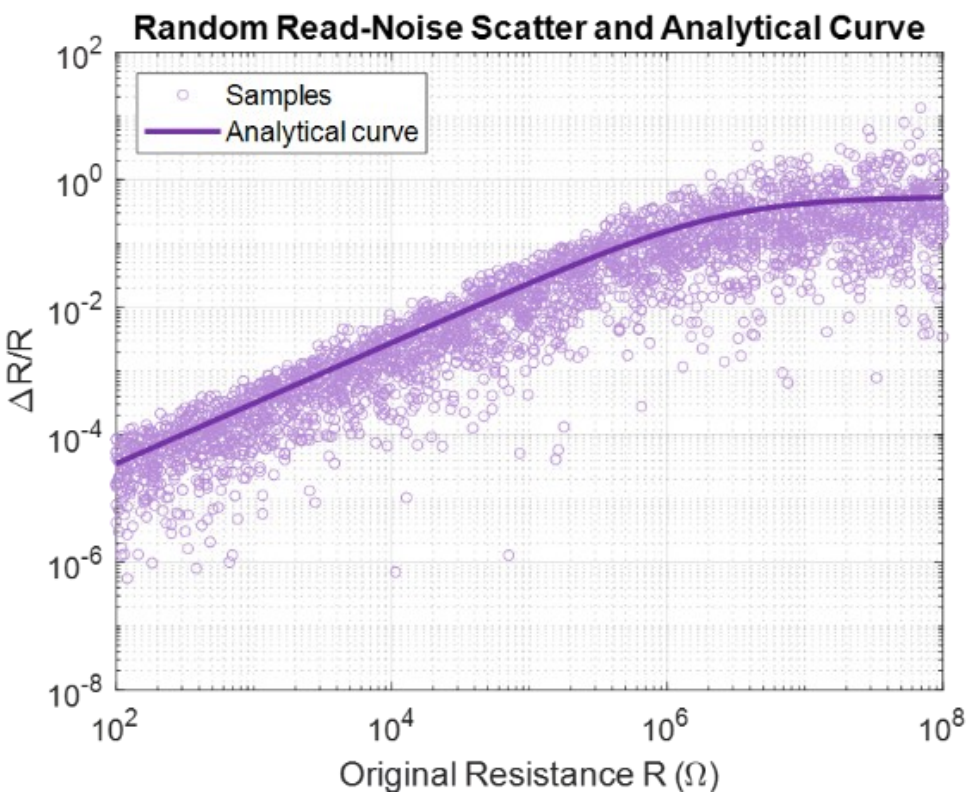

Fig. 11. Measured and analytically fitted relative read-noise variation versus resistance of the memristor devices.

Fig. 11 shows the statistical samples generated by the proposed model together with the fitted analytical curve of $\Delta R/R$ as a function of resistance. The results show that the relative fluctuation increases with resistance in the low- and intermediate-resistance regions and gradually approaches saturation at high resistance. This trend is consistent with the experimental observations reported for HfOx RRAM read noise, indicating that the proposed conductance-dependent model can capture the key state-dependent behavior required for crossbar- and system-level performance evaluation.

*3.4.3Write Scheme and Multi-Level Weight Realization*

The programmed conductance states in the proposed 1T1R synaptic array are obtained using a closed-loop write-verify scheme. Instead of relying on a single programming pulse, the target cell is iteratively updated through a sequence of write, read, and re-write operations. In each cycle, programming pulses with different amplitudes, widths, and pulse counts are applied according to the current conductance error. If the measured conductance does not reach the target level, additional write pulses are applied. If the cell is over-programmed, a reset operation is first performed, followed by a new write sequence. This process is repeated until the cell reaches the desired resistance range. Such a closed-loop strategy improves programming accuracy and enables reliable multi-level conductance tuning in the presence of device variability.

Fig. 12 illustrates the write operation for a selected target cell. As an example, the cell marked by the red star is programmed by applying a downward pulse sequence from 0.9 V to its corresponding column. At the same time, the gate-control line connected to the access transistor of the target branch, i.e., $S_{3_1}$, is biased by a 0.9 V source, while the associated row is maintained at 0.9 V. Under this bias condition, the selected memristor cell experiences the effective write voltage required for conductance update. To turn off unselected cells, all other gate-control lines are tied to ground. However, the cells sharing the same selected gate line $S_{3_1}$ may still be disturbed. To suppress unintended programming on these cells, all unselected columns are biased at 0.9 V, so that the corresponding transistors remain effectively turned off and no effective write voltage is developed across the unselected memristors.

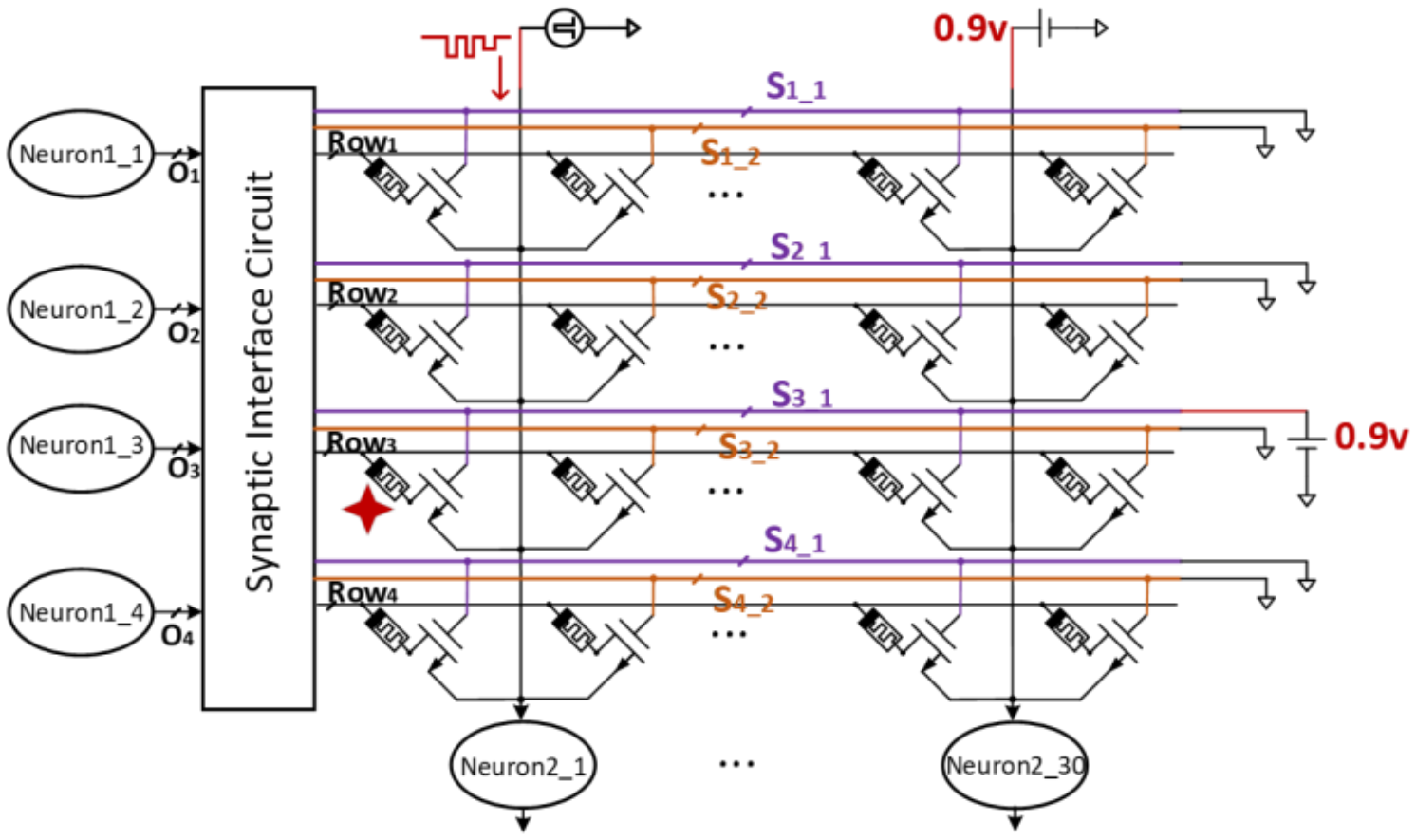

Fig. 12. Selective write scheme for the target memristive cell marked by the red star in the proposed 1T1R array.

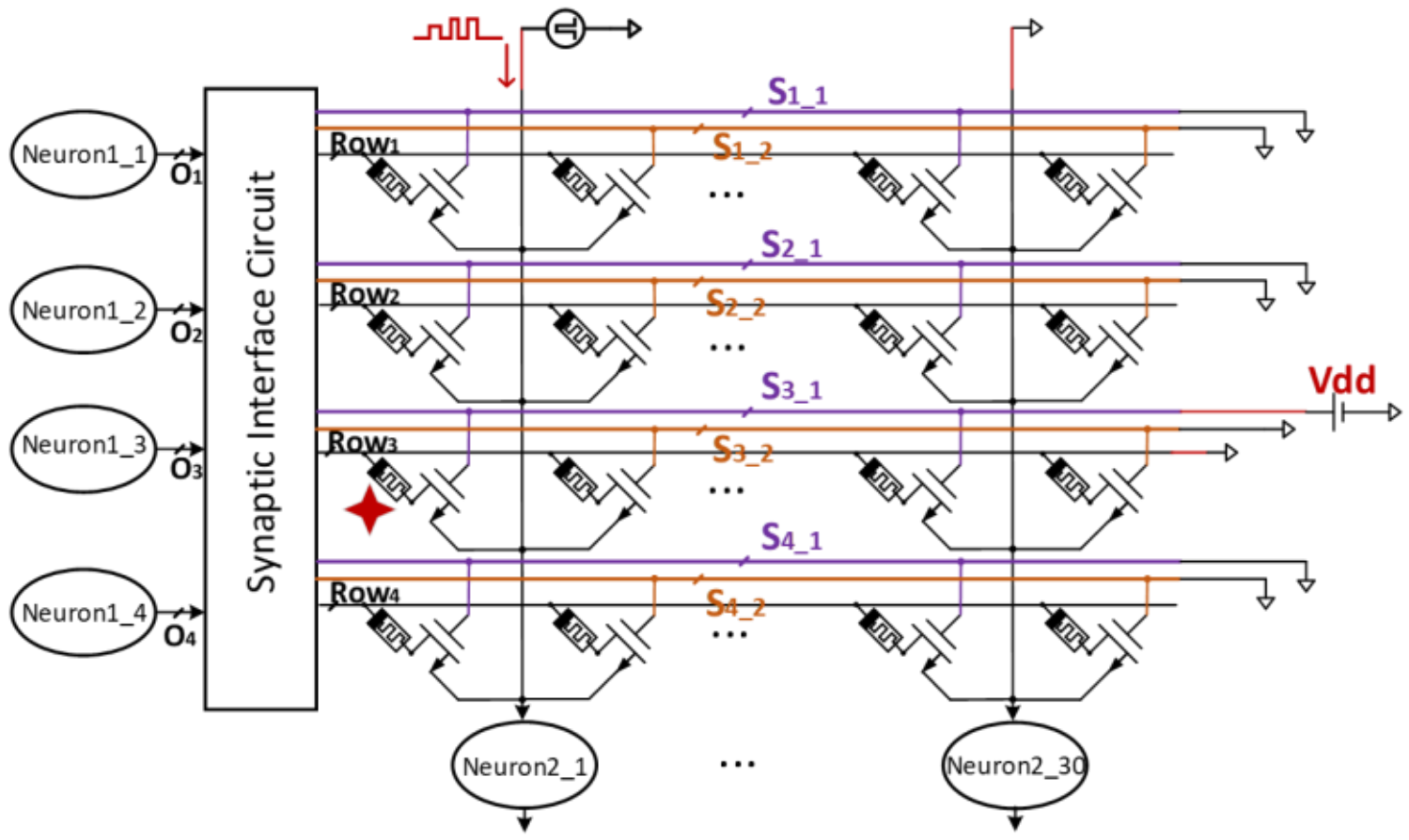

Fig. 13. Selective reset scheme for the target memristive cell marked by the red star in the proposed 1T1R array.

The corresponding reset operation is shown in Fig. 13. For the same target cell, the gate-control line $S_{3_1}$ is driven to Vdd, and the target column is excited by an upward pulse sequence starting from 0 V. Meanwhile, the row connected to the top electrode of the selected memristor, i.e., Row 3, is tied to ground. In this way, the target cell undergoes the reset bias required to increase its resistance state. To prevent disturbance to other cells sharing the same gate-control line, all other columns are also grounded during reset. Therefore, both write and reset operations are performed in a selective manner through coordinated control of the row, column, and gate-control signals.

Using this closed-loop programming framework, multiple resistance levels can be generated in the memristive synapse. However, due to the stochastic fluctuation observed during readout, the number of reliably distinguishable levels from a single memristor remains limited. In our measurements and post-processing, a single device can provide only 16 stable levels when read noise is taken into account. To further increase the effective synaptic resolution, we adopt a two-device parallel mapping scheme, where two memristors are used to represent one synaptic weight, analogous to a bit-slicing strategy. By combining the conductance contributions of two programmed devices, the number of available weight levels can be significantly enlarged. As shown in Fig. 14, this method yields 41 distinguishable levels, covering an equivalent resistance range from approximately 29 kΩ to 2 MΩ. This expanded level set provides a more flexible hardware mapping of synaptic weights under realistic device noise conditions.

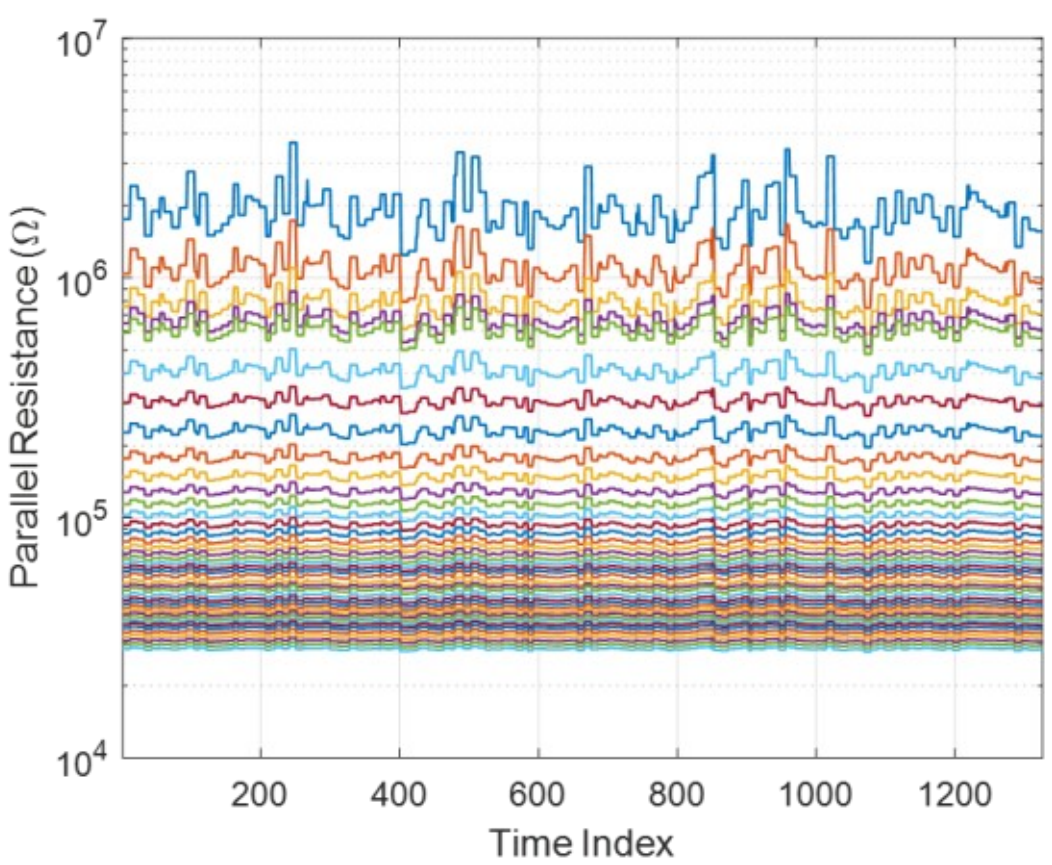

Fig. 14. 41 distinguishable synaptic levels realized by two-device parallel mapping under read-noise constraints.

## 4 SIMULATION AND RESULTS

In this section, we evaluate the proposed memristor-based SNN hardware from both circuit and system perspectives. We first describe the evaluation flow, including input encoding and hardware weight mapping from the trained SNN model to the memristive circuit implementation. Then, the circuit-to-system functionality is validated by comparing the hardware output with the SNNTorch reference and by demonstrating the closed-loop predator–prey tracking task. After that, the physical implementation and robustness of the proposed design are examined through layout results and Monte Carlo analysis. Finally, the energy and area of the proposed neuron are compared with prior works to highlight its hardware efficiency.

### 4.1 Evaluation Methodology and Hardware Mapping

To validate the proposed hardware framework, we establish a circuit-to-system evaluation flow that links software-trained SNN parameters to memristor-based hardware inference. The overall procedure consists of three major steps. First, the target network is trained in SNNTorch to obtain the synaptic weights and the corresponding reference outputs. Second, the trained weights are mapped to the memristive hardware according to the simulated electrical characteristics of the synaptic cells. Third, the mapped hardware model is used to perform inference, and the resulting outputs are compared with the software reference as well as the final task-level behavior.

#### *4.1.1Circuit-Level Input Encoding*

For input encoding, the normalized task variables are converted into spike-based stimuli and applied to the input neuron front end. Each input neuron is connected through an 800 kΩ input resistor, and a spike train with adjustable pulse interval is injected at the input. The neuron integrates the incoming pulses and produces output spikes. By sweeping the input spike interval over different values, we collect the corresponding output spike frequency of the neuron. This procedure establishes the mapping between the input spike frequency and the neuron output firing frequency, which is subsequently used to encode software-domain input features into hardware-operable spike signals for the synaptic array. For task-level hardware evaluation, the inference time of each frame is set to $24\ \mu s$.

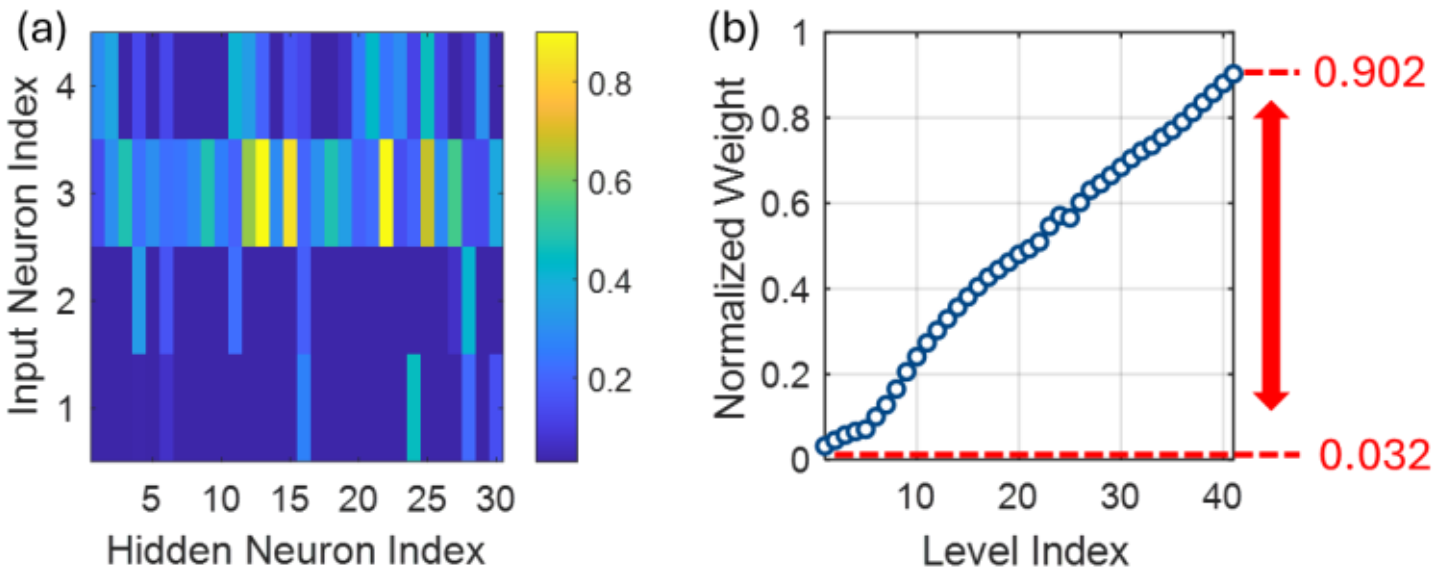

Fig. 15. Weight mapping from software to hardware: (a) SNNTorch-trained weight distribution; (b) hardware-mapped weight levels.

*4.1.2Circuit Level Weight Mapping*

Since each synaptic cell in the proposed memristive array provides only 41 distinguishable levels, the software-trained weights must be constrained to a hardware-realizable range before deployment. To minimize the loss of training effectiveness while maintaining compatibility with the limited cell resolution, we gradually reduced the training weight range and finally constrained it to 0.03-0.9, corresponding to a ratio of 30, as shown in Fig. 15(a).

To establish the relation between the memristor cell states and the effective synaptic weight in hardware, we sweep the programmable states of a target synaptic cell in the array. In the example shown in the figure, the characterized cell is the one highlighted by the shaded box. During this characterization, only the first input neuron is activated to generate output spikes, while the other input neurons remain in the resting state. For each cell level, a single input pulse is applied through the active input neuron, and the resulting membrane-voltage increment at the first neuron in the second layer is recorded. This allows us to extract the membrane-voltage contribution of each of the 41 realizable cell levels under one-pulse excitation. Since the effective integration range of the postsynaptic neuron is 0.21 V, the measured membrane-voltage change is normalized by 0.21 and used as the hardware-equivalent weight.

The resulting normalized values for all 41 levels are shown in Fig. 15(b). It can be observed that the mapped hardware weight range spans approximately from 0.032 to 0.902, which matches well with the software-side constrained range of 0.03-0.9. Based on this characterization, each trained software weight is assigned to the closest available hardware level, enabling direct deployment of the trained SNN parameters onto the memristive IMC array.

### 4.2 Circuit-to-System Validation Results

To validate the proposed memristor-based IMC hardware from circuit level to system level, we first examine the transient behavior of a representative inference sample. Fig. 16 shows the internal waveforms of the hardware inference process for one sample. The first four waveforms correspond to the spike inputs applied to the four input channels. The fifth waveform shows the membrane voltage of the first neuron in the hidden layer, where the incoming spikes are accumulated through the memristive synaptic array. Once the membrane voltage reaches the firing condition, the hidden neuron generates an output spike, as shown in the sixth waveform. This spike activity is then propagated to the output layer, where the seventh waveform shows the membrane-voltage evolution of the output neuron. Finally, the last waveform gives the corresponding output spike generated by the third-layer neuron. These results confirm that the proposed hardware correctly realizes the intended inference flow from input spike encoding, hidden-layer integration and firing, to output-layer accumulation and final spike generation.

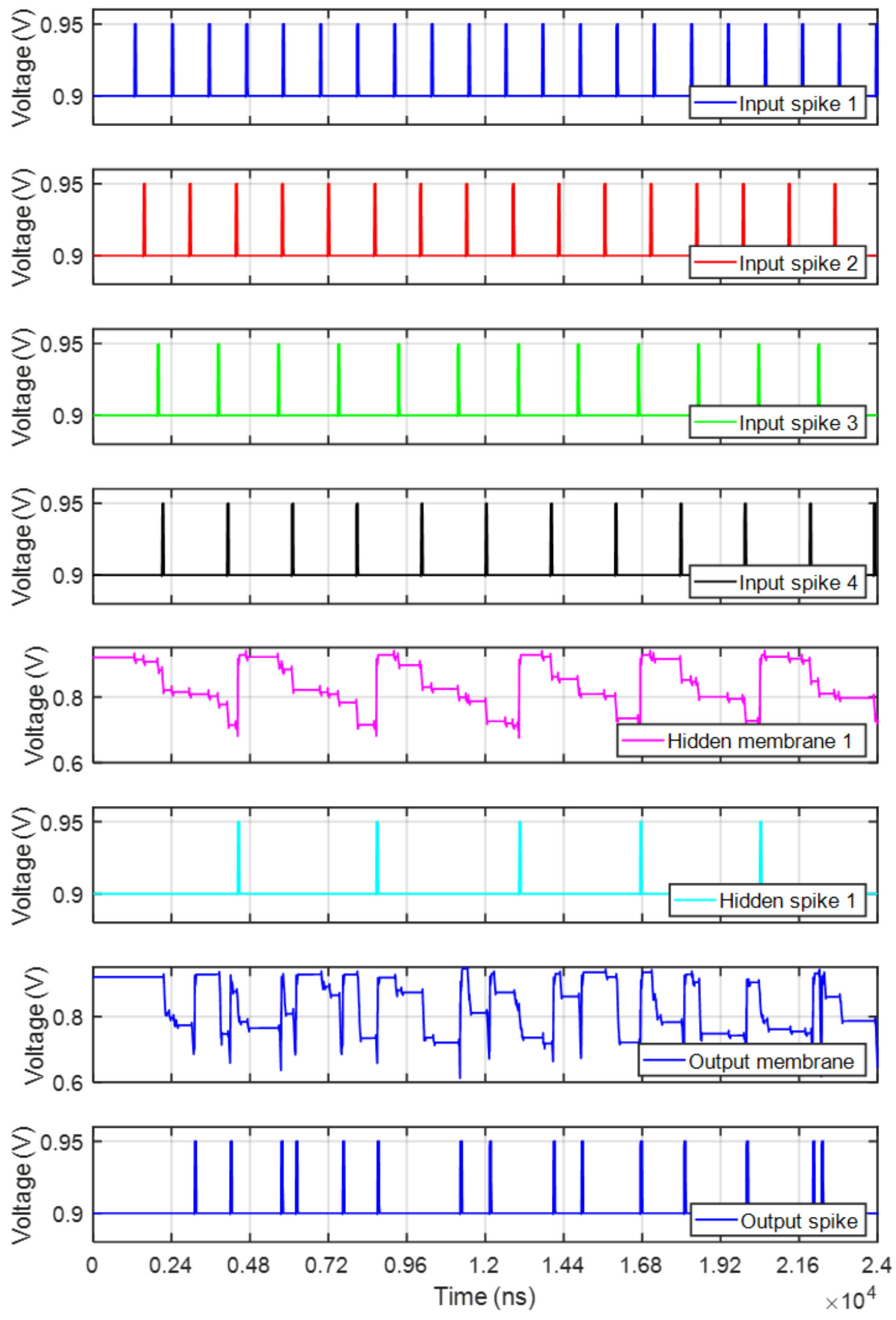

Fig. 16. Transient waveforms of a representative hardware inference sample.

After confirming the transient circuit behavior, we further evaluate the hardware output accuracy against the reference outputs generated by the trained SNNTorch model. A total of 659 samples from the inference dataset are applied to the hardware evaluation flow. For each sample, the normalized software input is converted into hardware spike-based stimuli, and the corresponding memristor-mapped circuit output is obtained through the proposed IMC framework. Fig. 17 compares the SNNTorch reference outputs with the memristor IMC circuit outputs. The results show strong agreement, with a correlation coefficient of 0.9622 and a mean absolute error (MAE) of 0.0544. These results indicate that the proposed hardware implementation can effectively reproduce the inference behavior of the trained SNN model, despite the limited device levels and non-ideal hardware mapping.

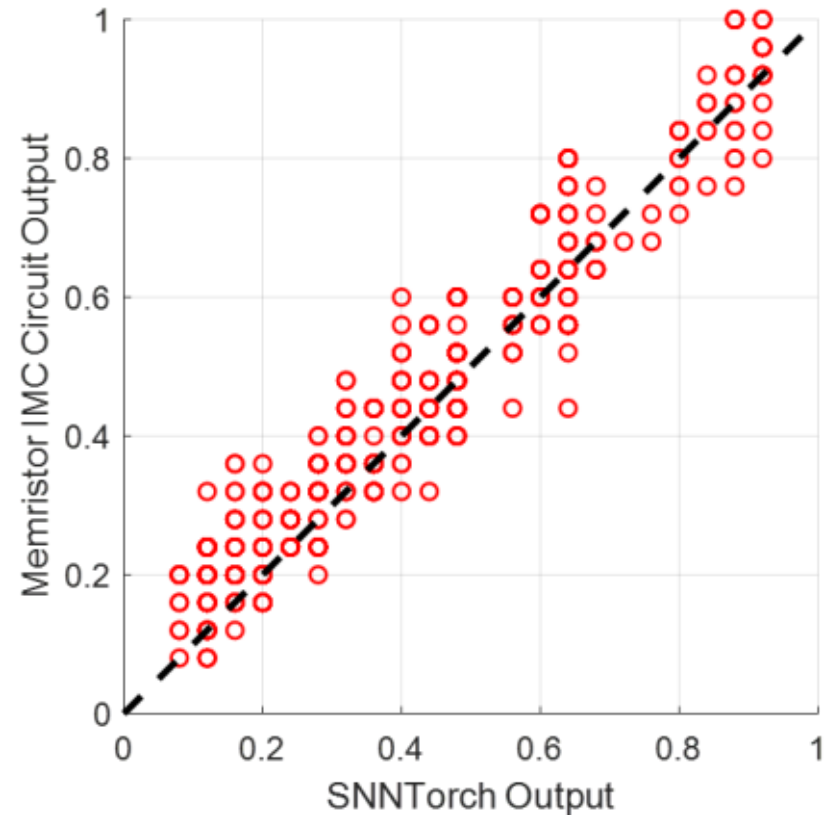

Fig. 17. Correlation between SNNTorch reference outputs and memristive IMC circuit outputs.

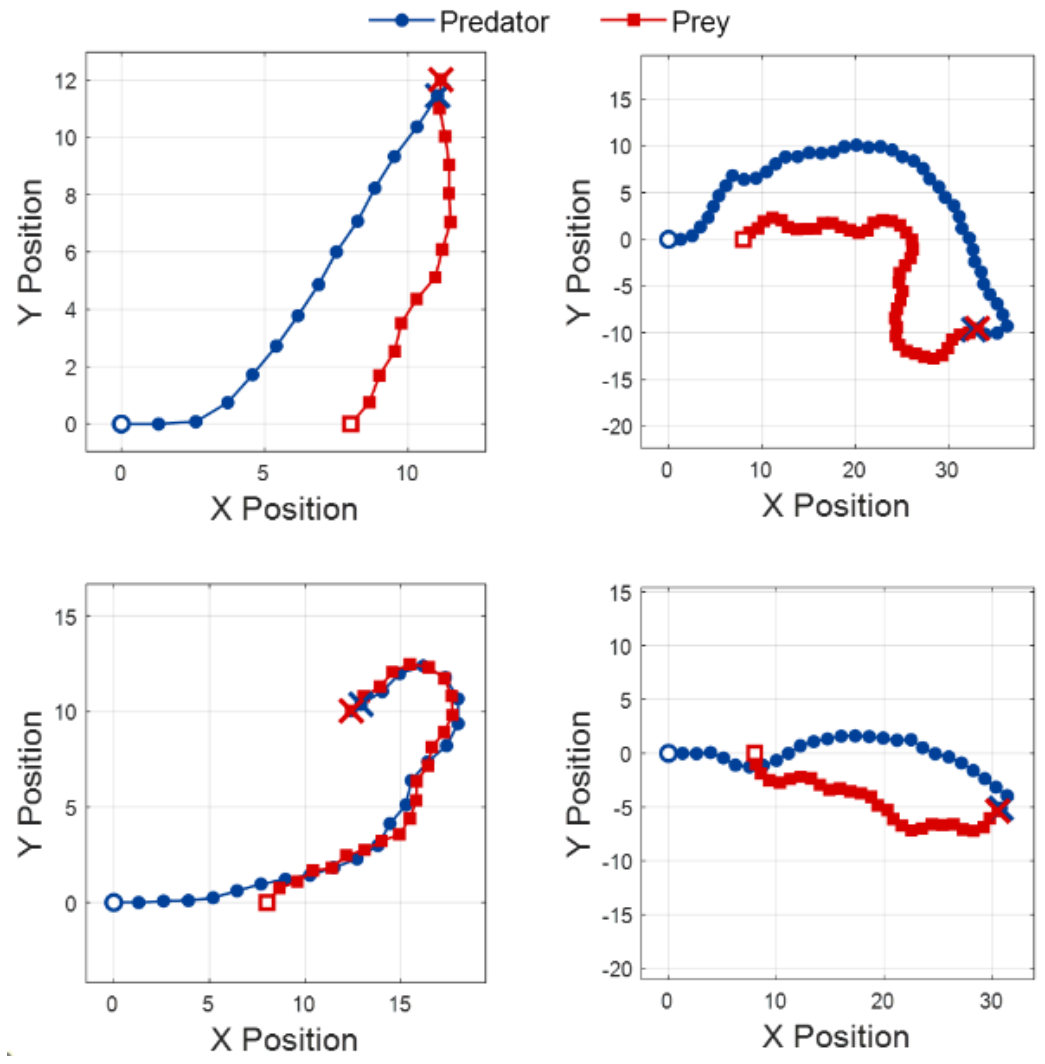

Fig. 18. Four representative closed-loop predator–prey trajectories generated using the proposed hardware inference.

To further evaluate the task-level effectiveness of the proposed hardware, the circuit-generated inference outputs were directly used in the closed-loop predator–prey tracking system. At each time step, the angular and positional information from the previous and current moments was taken as the input, encoded into spike-based hardware stimuli, processed by the memristor IMC circuit and then converted into the predator motion for the next step. A total of 500 complete trajectories were evaluated, and each trajectory was simulated for up to 50 time steps. A run was counted as successful only if the predator satisfied the capture condition within the 50-step window; otherwise, it is regarded as a failure. Under this criterion, 480 out of 500 trajectories were successfully captured, corresponding to a success rate of 96%. Representative examples are shown in Fig. 18, where the predator can progressively approach and capture the prey in most cases, demonstrating that the proposed hardware produces sufficiently reliable inference results to support closed-loop task execution at the system level.

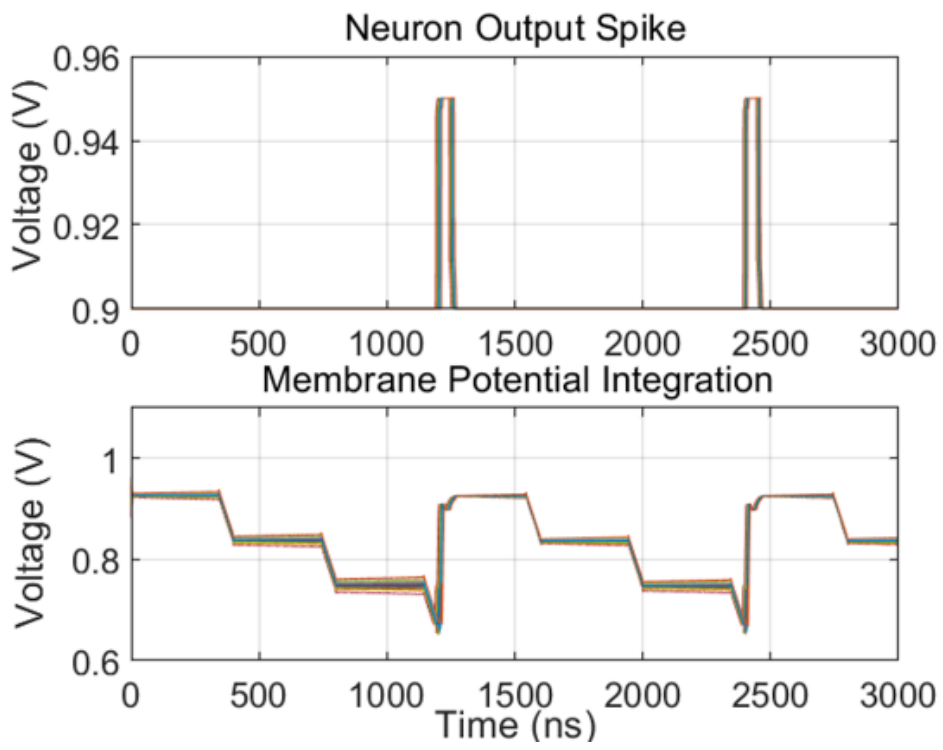

Fig. 19. Monte Carlo simulation results of the proposed neuron under device mismatch and process variations.

### 4.3 Robustness Analysis and Implementation Evaluation

To evaluate the robustness of the proposed design under circuit-level variations, we adopt a two-stage variability analysis flow. First, Monte Carlo simulations are performed on an isolated single neuron testbench to characterize the impact of both device mismatch and process-induced voltage/current variations on the neuron behavior. In this step, the neuron is excited by a pulse input, and the resulting membrane-voltage waveform and output spike waveform are recorded across multiple Monte Carlo runs. As shown in Fig. 19, these variations mainly introduce moderate spread in the membrane trajectory and small timing shifts in the output spike, while preserving the overall integrate-and-fire behavior.

The reason for first characterizing the isolated neuron is that the neuron constitutes the fundamental computational building block repeatedly used throughout the system. Direct transistor-level Monte Carlo simulation of the complete closed-loop memristive SNN system would be computationally prohibitive due to the large number of circuit instances and long transient simulation window. Therefore, instead of performing full-system Monte Carlo directly in Spectre, we used the neuron-level Monte Carlo samples as variation templates for the subsequent system-level robustness evaluation. In the system-level simulation, each neuron instance randomly draws one sample from the previously obtained Monte Carlo pool, thereby emulating a unique local variation condition for that neuron. Repeating this process for all neurons produces one complete variation realization of the entire system, and 200 such random realizations were evaluated in this work.

Using this hierarchical variability-injection approach, we further evaluated the closed-loop predator-prey tracking task at the system level. Representative results are shown in Fig. 20, where two example trajectory groups under variation are plotted. For each case, the predator trajectories generated from different random realizations remain closely clustered and preserve the same global pursuit trend, although limited path spread is still observed. This indicates that the neuron-level variations mainly perturb the detailed motion path without significantly altering the overall convergence behavior of the system. Therefore, the proposed design maintains stable task-level functionality under combined device mismatch and process-induced voltage/current variations, further supporting its robustness for system-level operation.

Fig. 21 presents the layout of the proposed neuron and synaptic interface circuitry. The implemented neuron occupies 906 μm2 and achieves an energy consumption of 10.67 pJ/spike. The area and energy are mainly contributed by the op-amp and comparator, which dominate the analog computation and output generation of the neuron. Table 1 summarizes the comparison between this work and prior designs [32-36]. It can be seen that the proposed design features relatively compact area and low energy consumption among the reported implementations.

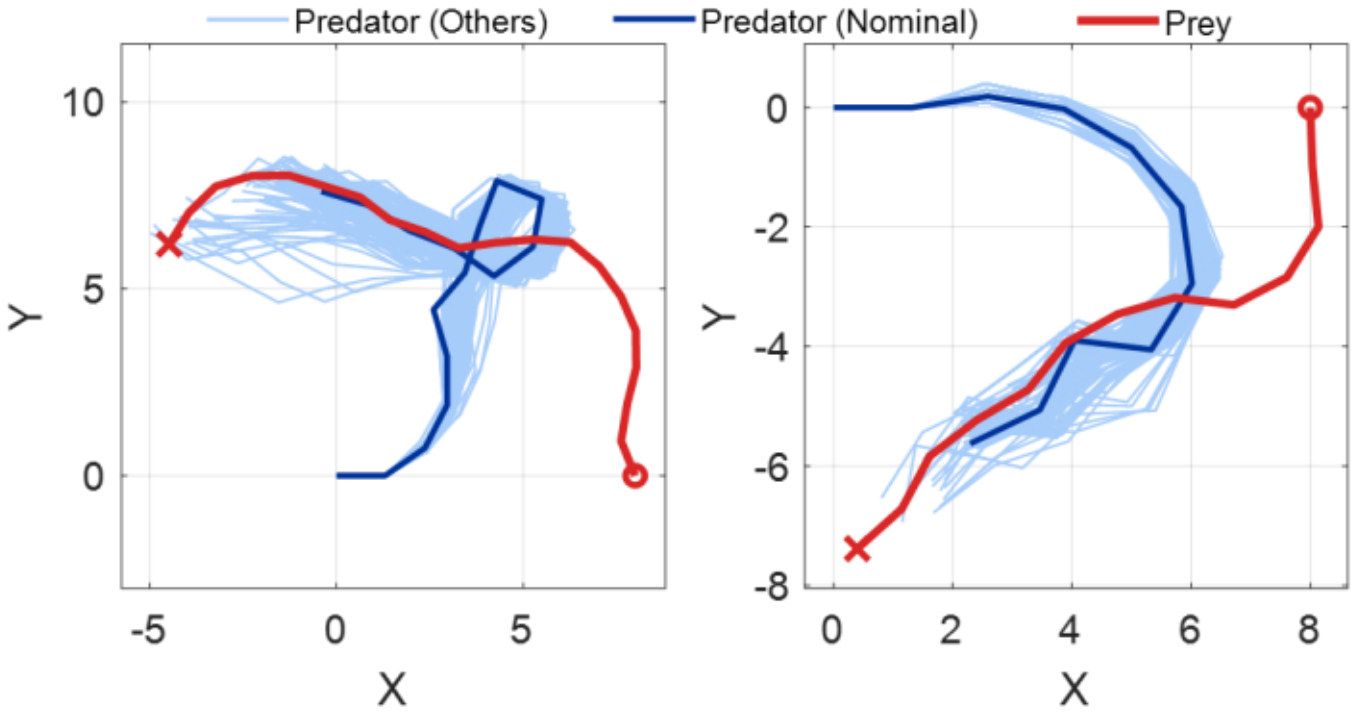

Fig. 20. Two representative closed-loop trajectory groups under neuron-level Monte Carlo variability injection.

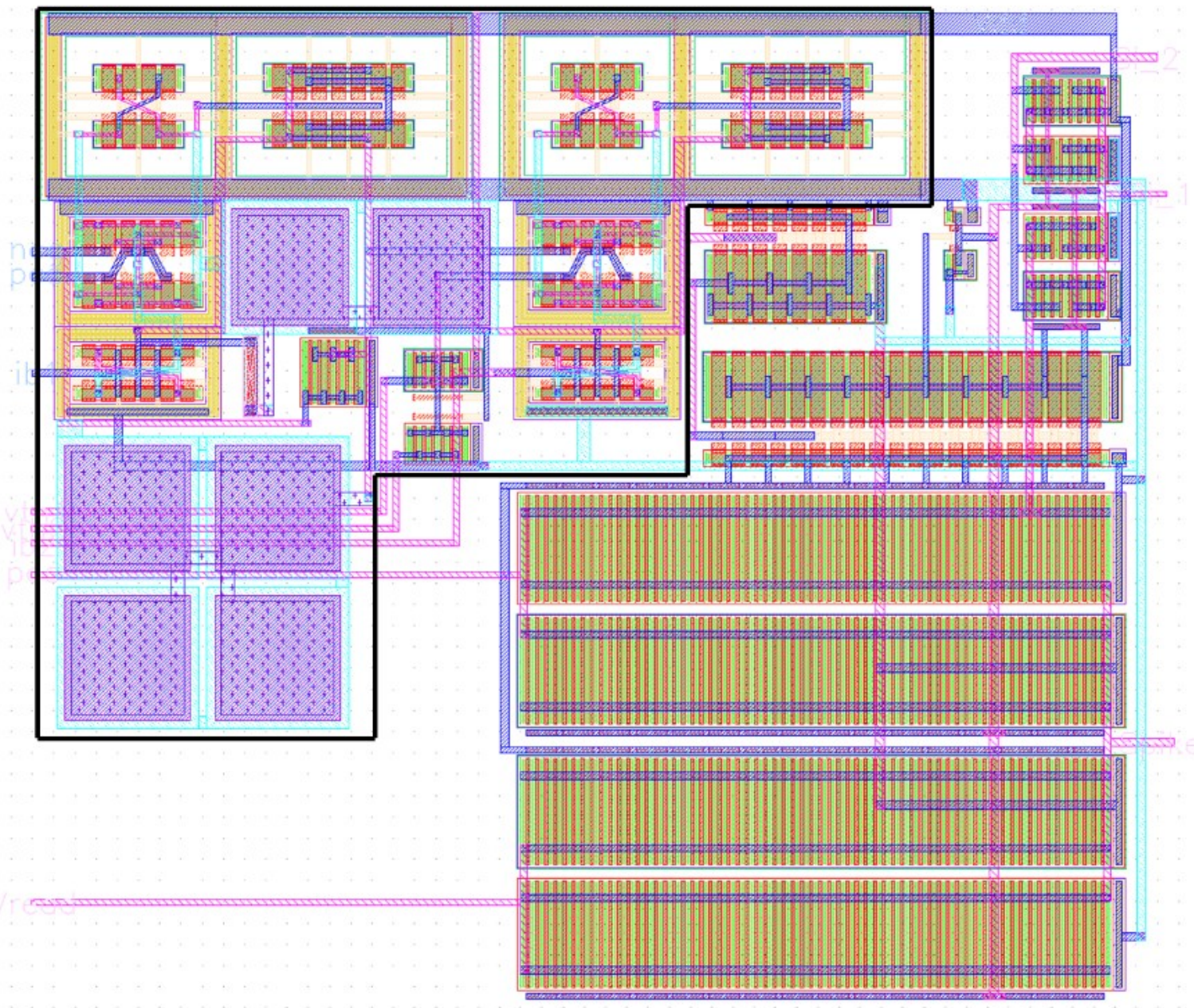

Fig. 21. Layout of the proposed neuron and synaptic interface circuitry, where the black box highlights the neuron core circuit.

## 5 CONCLUSION

In this paper, we presented a compact and energy-efficient memristive neuromorphic accelerator for bio-inspired interception tasks. The proposed architecture combines a 1T1R memristive crossbar, read-noise-aware hardware mapping, and an IF neuron with separated input and membrane nodes to enable reliable IMC-based SNN inference. Circuit- and system-level results show close agreement with the software SNN baseline, yielding a correlation coefficient of 0.9622 and a 96% interception success rate in the closed-loop predator-prey tracking task. Robustness analysis further confirms stable operation under device mismatch and process-induced voltage/current variations. Implemented in the SkyWater SKY130 PDK, the proposed neuron achieves 10.67 pJ/spike with an area of 906 μm². Together, these results position the proposed design as a compact, robust, and energy-efficient hardware solution for memristive SNN inference.

Table 1: Neuron performance comparison with prior works

| | [32] | [33] | [34] | [35] | [36] | This work |
|---|---|---|---|---|---|---|
| Implementation Style | analog | analog | analog | analog | mixed signal | analog |
| Technology node F (nm) | 130 | 65 | 180 | 22 | 65 | 130 |
| Supply Voltage (V) | 1.2 | 2.5/1.2 | 1.8 | 0.8 | 0.7-1.2 | 1.8 |
| Energy per spike per neuron (pJ) | 180 | 790 | 900 | 14 | 0.67 | 10.67 |
| Area per neuron ($\mu m^2$) | NA | 2352 | 2020 | 900 | 1410 | 906 |
| Normalized area ($\times 10^5 F^2$) | NA | 5.57 | 6.24 | 18.6 | 33.4 | 0.54 |

## REFERENCES

[1] Padamsey, Z. and Rochefort, N. L. Paying the brain's energy bill. Current opinion in neurobiology, 78 (2023), 102668.

[2] Ponulak, F. and Kasinski, A. Introduction to spiking neural networks: Information processing, learning and applications. Acta neurobiologiae experimentalis, 71, 4 (2011), 409-433.

[3] Tavanaei, A., Ghodrati, M., Kheradpisheh, S. R., Masquelier, T. and Maida, A. Deep learning in spiking neural networks. Neural networks, 111 (2019), 47-63.

[4] Fidjeland, A. K. and Shanahan, M. P. Accelerated simulation of spiking neural networks using GPUs. IEEE, City, 2010.

[5] Han, B. and Taha, T. M. Acceleration of spiking neural network based pattern recognition on NVIDIA graphics processors. Applied Optics, 49, 10 (2010), B83-B91.

[6] Bhuiyan, M. A., Pallipuram, V. K., Smith, M. C., Taha, T. and Jalasutram, R. Acceleration of spiking neural networks in emerging multi-core and GPU architectures. IEEE, City, 2010.

[7] Qu, Q., Lu, S., Shang, L., Jung, S., Liang, Q. and Pan, C. Fast and Energy-Efficient Analog Accelerator for Vision Transformer. City, 2025.

[8] Shang, L., Adil, M., Madani, R. and Pan, C. Fast Linear Programming Optimization Using Crossbar-Based Analog Accelerator. IEEE, City, 2020.

[9] Akopyan, F., Sawada, J., Cassidy, A., Alvarez-Icaza, R., Arthur, J., Merolla, P., Imam, N., Nakamura, Y., Datta, P. and Nam, G.-J. Truenorth: Design and tool flow of a 65 mw 1 million neuron programmable neurosynaptic chip. IEEE transactions on computer-aided design of integrated circuits and systems, 34, 10 (2015), 1537-1557.

[10] Chen, G. K., Kumar, R., Sumbul, H. E., Knag, P. C. and Krishnamurthy, R. K. A 4096-neuron 1M-synapse 3.8-pJ/SOP spiking neural network with on-chip STDP learning and sparse weights in 10-nm FinFET CMOS. IEEE Journal of Solid-State Circuits, 54, 4 (2018), 992-1002.

[11] Lu, S., Qu, Q., Jung, S., Liang, Q. and Pan, C. An Energy-Efficient RFET-Based Stochastic Computing Neural Network Accelerator. arXiv preprint arXiv:2512.22131 (2025).

[12] Zhang, X., Wang, W., Liu, Q., Zhao, X., Wei, J., Cao, R., Yao, Z., Zhu, X., Zhang, F. and Lv, H. An artificial neuron based on a threshold switching memristor. IEEE Electron Device Letters, 39, 2 (2017), 308-311.

[13] Yang, Z., Huang, Y., Zhu, J. and Ye, T. T. Analog circuit implementation of LIF and STDP models for spiking neural networks. City, 2020.

[14] Moriya, S., Yamamoto, H., Sato, S., Yuminaka, Y., Horio, Y. and Madrenas, J. A fully analog CMOS implementation of a two-variable spiking neuron in the subthreshold region and its network operation. IEEE, City, 2022.

[15] Wijesinghe, P., Ankit, A., Sengupta, A. and Roy, K. An all-memristor deep spiking neural computing system: A step toward realizing the low-power stochastic brain. IEEE Transactions on Emerging Topics in Computational Intelligence, 2, 5 (2018), 345-358.

[16] Joo, B., Han, J.-W. and Kong, B.-S. Energy-and area-efficient CMOS synapse and neuron for spiking neural networks with STDP learning. IEEE Transactions on Circuits and Systems I: Regular Papers, 69, 9 (2022), 3632-3642.

[17] Besrour, M., Lavoie, J., Omrani, T., Martin-Hardy, G., Koleibi, E. R., Menard, J., Koua, K., Marcoux, P., Boukadoum, M. and Fontaine, R. Analog Spiking Neuron in CMOS 28 nm Towards Large-Scale Neuromorphic Processors. arXiv preprint arXiv:2408.07734 (2024).

[18] Shooshtari, M., Serrano-Gotarredona, T. and Linares-Barranco, B. Review of Memristors for In-Memory Computing and Spiking Neural Networks. Advanced Intelligent Systems (2025), e202500806.

[19] Eshraghian, J. K., Wang, X. and Lu, W. D. Memristor-based binarized spiking neural networks: Challenges and applications. IEEE Nanotechnology Magazine, 16, 2 (2022), 14-23.

[20] Qu, Q., Lu, S., Shang, L., Jung, S., Liang, Q. and Pan, C. Variation-Aware Memristor-Based Analog Accelerator for Vision Transformer. Electronics, 15, 5 (2026), 1116.

[21] Yang, C., Wang, X. and Zeng, Z. Full-circuit implementation of transformer network based on memristor. IEEE Transactions on Circuits and Systems I: Regular Papers, 69, 4 (2022), 1395-1407.

[22] Dong, Z., Ji, X., Zhou, G., Gao, M. and Qi, D. Multimodal neuromorphic sensory-processing system with memristor circuits for smart home applications. IEEE Transactions on Industry Applications, 59, 1 (2022), 47-58.

[23] Plunkett, C. and Chance, F. Modeling Coordinate Transformations in the Dragonfly Nervous System. City, 2023.

[24] Eshraghian, J. K., Ward, M., Neftci, E. O., Wang, X., Lenz, G., Dwivedi, G., Bennamoun, M., Jeong, D. S. and Lu, W. D. Training spiking neural networks using lessons from deep learning. Proceedings of the IEEE, 111, 9 (2023), 1016-1054.

[25] Venker, J. S., Vincent, L. and Dix, J. A low-power analog cell for implementing spiking neural networks in 65 nm CMOS. Journal of Low Power Electronics and Applications, 13, 4 (2023), 55.

[26] Soupizet, T., Jouni, Z., Wang, S., Benlarbi-Delai, A. and Ferreira, P. M. Analog spiking neural network synthesis for the MNIST. Journal of Integrated Circuits and Systems, 18, 1 (2023), 1-12.

[27] Chua, L. Memristor-the missing circuit element. IEEE Transactions on circuit theory, 18, 5 (2003), 507-519.

[28] Strukov, D. B., Snider, G. S., Stewart, D. R. and Williams, R. S. The missing memristor found. nature, 453, 7191 (2008), 80-83.

[29] Li, H., Wu, T. F., Mitra, S. and Wong, H.-S. P. Resistive RAM-centric computing: Design and modeling methodology. IEEE Transactions on Circuits and Systems I: Regular Papers, 64, 9 (2017), 2263-2273.

[30] Li, H., Jiang, Z., Huang, P., Wu, Y., Chen, H.-Y., Gao, B., Liu, X., Kang, J. and Wong, H.-S. Variation-aware, reliability-emphasized design and optimization of RRAM using SPICE model. IEEE, City, 2015.

[31] Ambrogio, S., Balatti, S., Cubeta, A., Calderoni, A., Ramaswamy, N. and Ielmini, D. Statistical fluctuations in HfO x resistive-switching memory: Part II—Random telegraph noise. IEEE Transactions on Electron Devices, 61, 8 (2014), 2920-2927.

[32] Valentian, A., Rummens, F., Vianello, E., Mesquida, T., de Boissac, C. L.-M., Bichler, O. and Reita, C. Fully integrated spiking neural network with analog neurons and RRAM synapses. IEEE, City, 2019.

[33] Aamir, S. A., Stradmann, Y., Müller, P., Pehle, C., Hartel, A., Grübl, A., Schemmel, J. and Meier, K. An accelerated LIF neuronal network array for a large-scale mixed-signal neuromorphic architecture. IEEE Transactions on Circuits and Systems I: Regular Papers, 65, 12 (2018), 4299-4312.

[34] Asghar, M. S., Arslan, S. and Kim, H. A low-power spiking neural network chip based on a compact LIF neuron and binary exponential charge injector synapse circuits. Sensors, 21, 13 (2021), 4462.

[35] Rubino, A., Livanelioglu, C., Qiao, N., Payvand, M. and Indiveri, G. Ultra-low-power FDSOI neural circuits for extreme-edge neuromorphic intelligence. IEEE Transactions on Circuits and Systems I: Regular Papers, 68, 1 (2020), 45-56.

[36] Ko, Y., Kim, S., Shin, K., Park, Y., Kim, S. and Jeon, D. A 65 nm 12.92-nj/inference mixed-signal neuromorphic processor for image classification. IEEE Transactions on Circuits and Systems II: Express Briefs, 70, 8 (2023), 2804-2808.